\documentclass[aps,prb,10pt,twocolumn,showpacs,amsmath,amssymb,superscriptaddress]{revtex4-1}
\usepackage{graphicx}
\usepackage{color}
\usepackage{float}
\usepackage{bm}

\newcommand{\beq}{\begin{equation}}
\newcommand{\eeq}{\end{equation}}

\newcommand{\beqa}{\begin{eqnarray}}
\newcommand{\eeqa}{\end{eqnarray}}

\newcommand{\etal}{\mbox{\textit{et al.}}~}

\newcommand{\Eqref}[1]{Eq.~(\ref{#1})}
\newcommand{\Figref}[1]{Fig.~\ref{#1}}

\newcommand{\ie}{{\em i.\,e. }}

\newcommand{\BK}[1][]{\mathbf{K}_{#1}}

\newcommand{\BD}{\boldsymbol{D}}

\newcommand{\BPi}[1][]{\boldsymbol{\Pi}_{#1}}

\newcommand{\Hi}[1 ]{\mathcal{H}\left[#1\right]}

\DeclareMathOperator{\Tr}{Tr}
\newcommand{\Trr}[1 ]{\Tr \left[#1\right]}
\newcommand{\Ftw}[1]{\int \frac{d\omega}{2\pi}e^{-i\omega (t-t')}#1}

\renewcommand{\Re}{\operatorname{Re}}
\renewcommand{\Im}{\operatorname{Im}}
\def\5!{\!\!\!\!\!}

\begin{document}

\title{Current-induced runaway vibrations in dehydrogenated graphene nanoribbons}
\author{Rasmus Bjerregaard Christensen}
\affiliation{Center for Nanostructured Graphene (CNG), Department of Micro- and Nanotechnology, Technical University of Denmark, {\O}rsteds Plads, Bldg. 345E, DK-2800 Kongens Lyngby, Denmark}

\author{Jing-Tao L\"u}
\affiliation{School of Physics, Huazhong University of Science and Technology, 430074 Wuhan, P. R. China }

\author{Per Hedeg\aa rd}
\affiliation{Niels-Bohr Institute and Nano-Science Center, University of Copenhagen, Universitetsparken 5, 2100 Copenhagen \O, Denmark}

\author{Mads Brandbyge}
\email{mads.brandbyge@nanotech.dtu.dk}
\affiliation{Center for Nanostructured Graphene (CNG), Department of Micro- and Nanotechnology, Technical University of Denmark, {\O}rsteds Plads, Bldg. 345E, DK-2800 Kongens Lyngby, Denmark}
\begin{abstract}
We employ a semi-classical Langevin approach to study
current-induced atomic dynamics in a partially dehydrogenated armchair graphene
nanoribbon. All parameters are obtained from density functional theory.
The dehydrogenated carbon dimers behave as effective impurities,
whose motion decouples from the rest of carbon atoms. The electrical
current can couple the dimer motion in a coherent fashion. The coupling, which is mediated by 
nonconservative and pseudo-magnetic current-induced forces, change the atomic dynamics, and thereby show 
their signature in this simple system. We study the atomic dynamics and current-induced vibrational instabilities using a simplified eigen-mode analysis.  
Our study shows that the armchair nanoribbon serves as a possible testbed for probing the current-induced forces.
\end{abstract}

\maketitle

\section{Introduction}
The electronic and transport properties of graphene has been the focus of
intense study since its discovery in $2004$\cite{Neto2009}.  Due to the strong
$\sigma$-bonding between carbon atoms, graphene has a very high thermal
conductivity, and can potentially sustain much higher current intensity than
other materials. Graphene nanoribbons are pesent a bandgap
opening due to quantum confinement in the transverse ribbon direction.  This
opens the possibilities of realizing various electronic devices, especially
field-effect transistors, using graphene nanoribbons. Atomically precise ribbons\cite{Cai2010}, as well as more 
advanced ribbon-based structures\cite{Cai2014,Liu2015},
has been fabricated "bottom-up" on metal surfaces. Through-ribbon conductance has been investigated using STM\cite{Koch2012}, and the electron-vibration signals in the
current has been addressed by theory\cite{Christensen2015}.

When cutting into one-dimensional ribbons, dangling bonds emerge at the
boundary carbon atoms. If there is an electrical current passing through the
ribbon, we expect these boundary atoms with dangling bonds are mechanically
weak compared to the central atoms. Actually, it has been observed
experimentally that these atoms can be removed by the passing electrical
current due to current-induced local
heating\cite{JiaX2009sci,EngelundM2010prl}. One theoretical study suggest that
the carbon dimers at the armchair edge vibrates locally and interacts
strongly with the electrical current\cite{EngelundM2010prl}.  They can be thought as atomic scale
defects at the boundary. How the current-induced forces affect the dynamics
of these dimers is an interesting question to ask since it could be addressed by experiments.  
Employing a semi-classical Langevin approach, we have previously studied the current-induced atomic
dynamics a graphene nano-constriction\cite{Gunst2013}.  
However, the number of atoms involved even in such a small system makes a simple analysis difficult.

On the other hand, since the first prediction of energy non-conservative nature of
current-induced forces,\cite{DuMcTo.2009} there has been a substantial theoretical
effort aimed at exploring its consequences for the stability of current-carrying nano-systems\cite{Lu2010e,luprb12,Dundas2012,Cunningham2014,Todorov2014}, 
or the possibility of driving atomic motors\cite{DuMcTo.2009,Bustos2013}.  Moreover, it has been shown that, in
addition to the non-conservative force, the current-induced forces also include
an effective Lorentz force or pseudo-magnetic force, originated from the Berry phase of
electrons\cite{Lu2010e}. Performing similar analysis using a scattering theory
approach shows that the predictions apply equally well to much larger mesoscopic
coherent conductors\cite{BoKuEgVo.2011,Bustos2013}. A requirement of impact of the non-conservative force is that two or more vibrational modes close in 
frequency couple to each other via the current carrying electronic states. This can establish a generalized circular "water-wheel" motion, either in real-space\cite{DuMcTo.2009} or in mode-space.
Another requirement is that these modes have little damping due to the coupling to the phonon reservoir. Unfortunately, there has
not been a clear experimental setup where these new theoretical findings can be put to a test proving their effect in an unambiguous way. 
Thus, it is of interest to be able to propose such a setup based on first principles calculations with realistic unadjustable parameters.

In this paper, we study the current-induced dynamics in a partially
dehydrogenated armchair graphene ribbon. We show that, atomic motion of the
dehydrogenated carbon dimer at the nanoribbon boundaries are relatively
decoupled from other dimers and also the rest carbon atoms. This results in
several nearly degenerate atomic vibrations, where each of these involves mainly one
dimer. However, a coupling of the dimer vibrations takes place via the flowing electrical
current. All these features are favorable to observe the effect of
current-induced forces, thus making armchair nanoribbon as an ideal candidate
to study.

In the rest of the paper, we summarize our theoretical (Sec.~\ref{sec:theory})
and numerical (Sec.~\ref{sec:num}) methods, and present our analysis of the
armchair graphene ribbon (Sec.~\ref{sec:results}).  We end this paper with 
our concluding remarks (Sec.~\ref{sec:conclusion}).
\section{Theory}
\label{sec:theory}
\begin{figure}[t]
   \includegraphics[scale=1]{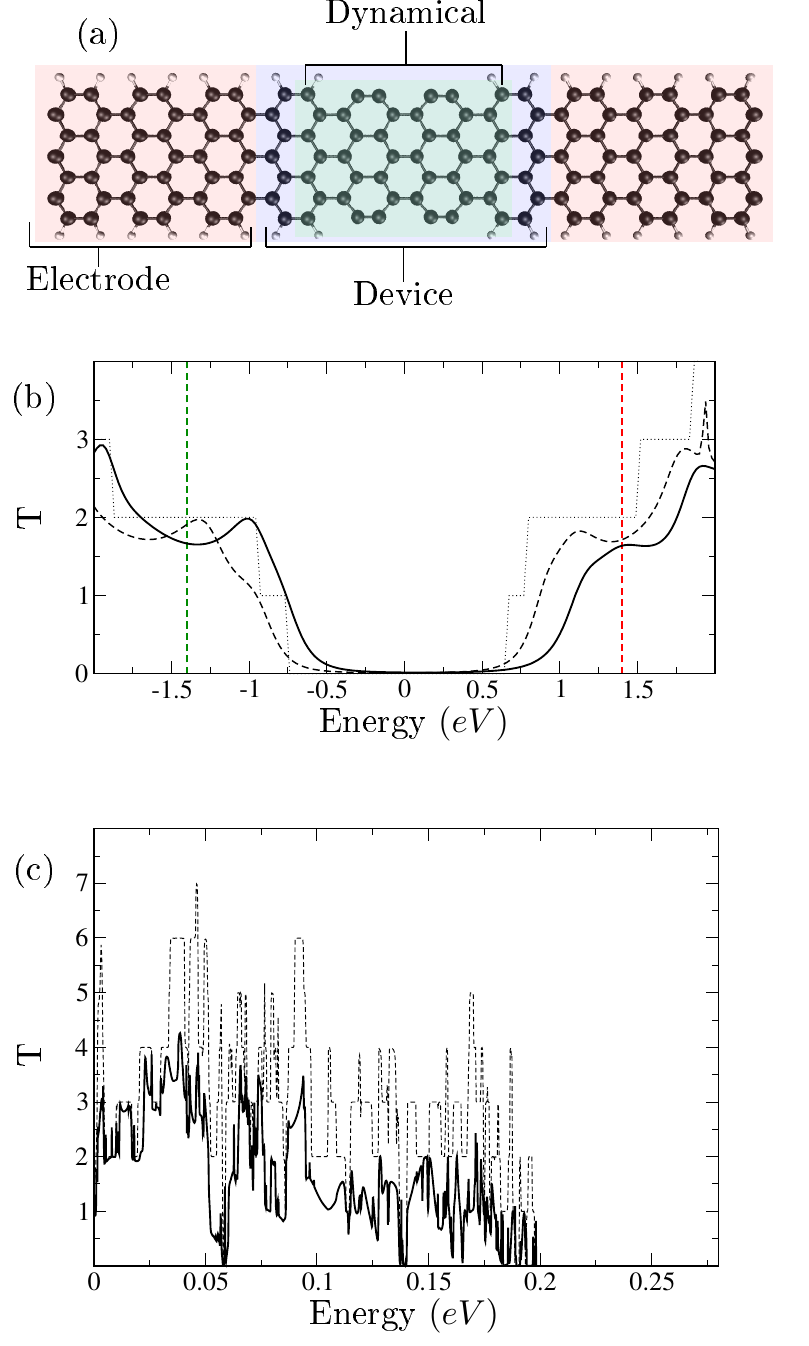}
   \caption{(a) Structure of the transport setup defining device and symmetric electrode(left shown) regions. The motion of atoms is considered in the dynamical region. 
   	(b) Electron transmission in a perfect, infinite ribbon (dotted), and  
   	with broadened states in the electrodes to mimic metallic contacts with/without dehydrogenation (dashed/full). Red and green 
   	vertical dashed lines indicate the shifts in Fermi energy used in \Figref{fig:DOS}). (c) Solid line is phonon
   transmission for the structure in (a), dotted line is phonon transmission for a
 pristine hydrogenated ribbon.}
   \label{fig:Setup}
\end{figure}

We consider a standard Landauer-type transport setup described in \Figref{fig:Setup}(a). The
system/device is in contact with the left and right leads. Each lead serves as
both electronic and phononic bath. The bath degrees of freedom are
non-interacting. We are interested in the atomic dynamics in the device region (displacements $U$),
which can be described by the semi-classical generalized Langevin
equation(SGLE)\cite{luprb12,FEYNMAN1963,CALE.1983,SC.1982},
\begin{equation}
	\ddot U(t)-F(U(t)) = -\int^t {\Pi^r}(t-t')U(t')dt' + f(t).
	\label{eq:langa}
\end{equation}
Here $F(U)$ is the force between atoms in the device region and $f(t)$ is a random force due to  
thermal and voltage-bias-induced fluctuations.  
The $\Pi^r$ (self-energy) describes the time-delayed back action of the bath on the system due to the motion of the system. It has three contributions,
\begin{equation}
 \Pi^r=\Pi^r_L +\Pi^r_R+\Pi^r_e,
\end{equation}
where $\Pi^r_L$ and $\Pi^r_R$ describe the coupling to the phonon reservoirs outside device, while the self-energy $\Pi^r_e$
describes the coupling to the electrons. In equilibrium, this result has been
obtained by f.ex. Head-Gordon and Tully\cite{HETU.1995}. In principle we may apply the SGLE including the non-linear part of $F$ \cite{Lu2011},
but here we will restrict ourselves to the harmonic approximation, $F(U)=-{\bf K}U$.

\subsubsection{Non-equilibrium}
The semi-classical Langevin equation can be extended to include the non-equilibrium effects in
the electronic system due to the current\cite{BrHe.1994a,Lu2010e,luprb12}. 
In accordance with intuition the "traditional" Joule-heating is present in the fluctuating force, $f$, while the current-induced forces show up in $\Pi^r_e$. Here we focus on the latter.
The contribution from the electron degrees of freedom including the non-equilibrium effects can be expressed in terms of the
coupling-weighted electron-hole pair density of states, $\Lambda$,
\begin{align}
\Pi^r_e(t-t')=&\theta(t-t')\Ftw{\Lambda(\omega)}\,
\end{align}
with $\Lambda$ (including spin), given by,
\begin{align}
\Lambda(\omega) =& \sum_{\alpha,\beta}\Lambda^{\alpha \beta}(\omega),
\\ 
\Lambda_{kl}^{\alpha \beta}(\omega)&=2\int\!\!\frac{d\epsilon_{1}}{2\pi}\int\!\!\frac{d\epsilon_{2}}{2\pi}\delta(\epsilon_{1}-\epsilon_{2}-\hbar\omega)\nonumber\\
&\times\Tr\left[M^{k}A_{\alpha}(\epsilon_{1})M^{l}A_{\beta}(\epsilon_{2})\right]\nonumber\\
&\times \left(n_{F}(\epsilon_{1}-\mu_{\alpha})-n_{F}(\epsilon_{2}-\mu_{\beta})\right) \label{eq:lambda}.
\end{align} 
Here, $A_{\alpha/\beta}$ is the density of scattering states incoming from left and right electrodes (indices $\alpha$ and $\beta$), 
while $M$ is electron-phonon couplings ($k$ and $l$ phonon indices). One can loosely think of the motion of phonon $k$ excites an electron-hole pair of 
energy $\hbar\omega$ which is absorbed by phonon $l$.

The SGLE, in \Eqref{eq:langa}, is given in the time domain. However, since we
are considering steady state, it is convenient to  work in the frequency
domain.  Thus, by Fourier transformation we obtain,
\begin{align}
	\Pi^r_e(\omega) = \int_{-\infty}^\infty{d\omega'}\frac{\Lambda(\omega')}{\omega'-\omega-i\eta} 
\end{align} 
By applying the Sokhatsky-Weiestrass theorem $\Pi^r(\omega)$ can be split into four contributions giving rise to the four forces
\begin{eqnarray}
\Pi_e^r(\omega) &=  \underbrace{i\pi \Re\Lambda(\omega)}_{\rm FR}-\underbrace{\pi \Im\Lambda(\omega)}_{\rm NC}  \nonumber\\
&+\underbrace{\pi\Hi{\Re\Lambda(\omega')}(\omega)}_{\rm RN}+\underbrace{i\pi\Hi{\Im\Lambda(\omega')}(\omega)}_{\rm BP}.
\end{eqnarray}
Here, FR, NC, RN, BP represent the electronic friction, non-conservative
force, renormalization of the atomic potential, and Berry-phase-induced
pseudo-magnetic force, respectively\cite{luprb12}. 

\subsubsection{Run-away modes}
In order to analyze the influence of the current we define the nonequilibrium
phonon density of states as,
\begin{equation}
  {\rm DOS}(\omega)= -\frac{\omega}{2\pi}\Im\Trr{\BD^r(\omega)},
\label{eq:PHDOS}
\end{equation}
where $\BD^r(\omega)$ is the nonequilibrium phonon Greens function obtained from the SCLE,
\begin{equation}
\BD^r(\omega)= \frac{1}{(\omega+i\eta)^2-\BK-\BPi^r(\omega)}.
\end{equation}
Note that we introduced bold-face to underline that these are matrices with mode-index ($k,l$).
Contrary to the equilibrium situation, the DOS given in \Eqref{eq:PHDOS} can
take negative values at certain peak values, due to the electronic current.  We can interpret a negative
peak in the DOS at a frequency $\omega_0$ as modes at $\omega_0$ with a negative lifetime, \ie with growing in amplitude
as a function of time and denote these by "run-away" modes.

\subsubsection{Mode analysis}
In order to identify the modes which can show run-away behavior, we need to find the
solutions to  \Eqref{eq:langa}, setting the driving noise force,
$f(\omega)$, to zero.  This is done by  treating the velocity and displacement
as independent variables and use the relation $-i\omega
U(\omega)=\dot{U}(\omega)$ to obtain the double-sized eigenvalue problem. However, the self-energy $\BPi^r$ is frequency-dependent. Thus, to analyze a specific runaway mode giving rise to a negative peak in \Figref{fig:DOS} (a), we evaluate the self-energy at the negative peak frequency $\omega_0$,  

\begin{widetext}
\begin{equation}
\begin{bmatrix}
\bf{0} & -\bf{1}
\\
\BK + \Re[\BPi[ph]^r(\omega_0)+\BPi[e]^r(\omega_0)]  & -\Im[\BPi[ph]^r(\omega_0)+\BPi[e]^r(\omega_0)]/\omega_0  
\end{bmatrix}
\begin{bmatrix}  
U(\omega)
\\
\dot{U}(\omega)
\end{bmatrix}
= i\omega 
\begin{bmatrix}
U(\omega)
\\
\dot{U}(\omega)
\end{bmatrix}.
\label{eq:WBAdouble}
\end{equation}
\end{widetext}

Thus, the dynamical matrix is renormalized by $\Re[\BPi[ph]^r(\omega_0)+\BPi[e]^r(\omega_0)]$
and the friction originates from  $\Im[\BPi[ph]^r(\omega_0)+\BPi[e]^r(\omega_0)]/\omega_0$. 
Solving \Eqref{eq:WBAdouble} gives a set of eigenmodes and complex eigenfrequencies, but only the "self-consistent" 
mode, which fulfills $\Re\omega=\omega_0$, is relevant.

For a given eigenmode a corresponding positive imaginary part of the
eigenfrequency designates that the mode is  a run-away mode, while if the imaginary
part is negative the mode is damped. The damping can be quantified by the
inverse $Q$-factor giving the change in energy per period 
\begin{equation}
{Q_i}^{-1}=\left(2\pi\frac{\Delta E_i}{E_{i,tot}}\right)^{-1}=-\frac{2\Re\omega_i}{\Im\omega_i}\,.       
\end{equation}
Thus, the run-away modes can be identified as the modes where $\Im\omega>0$.
The run-away modes are a linear combination of the non-perturbed normal modes.
Normally, the runaway makes closed loops in real or in abstract mode space. Thus,
the NC force allows the mode to pick up energy every time a loop is completed,
eventually leading to break down of the harmonic approximation, ending with e.g.
rupture or damping by anharmonic effects leading to a limit cycle motion\cite{BoKuEg.2012}.

\section{Numerical calculation}
\label{sec:num}
We have calculated the electronic and phononic structure of the graphene nanoribbon
from density function theory (DFT) using the SIESTA/TranSIESTA
codes\cite{Soler.02,BrMoOr.2002}. 
The generalized gradient approximation is used for the
exchange-correlation functional, and a single-$\zeta$ polarized basis set is
used for the carbon and hydrogen atoms.  A cut-off energy of $400$ Ry
is used for the real-space grid.  The electron-vibrational coupling is
calculated using the INELASTICA package, which uses a finite difference
scheme\cite{FrPaBr.2007}.

\section{Results}
\label{sec:results}
The partial dehydrogenated graphene nanoribbon we considered is shown in
\Figref{fig:Setup}(a), where four hydrogen atoms have been removed on each
side of the ribbon. In principle, dehydrogenation could be performed at chosen positions with a STM\cite{Wang2010}.
The same structure has been considered in our recent work,
focusing on the asymmetry in phonon emission and heat distribution due to the nonequilibrium for lower voltage than where run-away instability occurs\cite{Lu2015}. This asymmetry is intrinsically linked to momentum transfer from electrons to phonons and thus to the non-conservative current-induced forces ("electron-wind")\cite{Todorov2011}. 
As we mentioned before, the reason we choose this structure is that, the dehydrogenated carbon dimers
can be considered as ``defects'' in the armchair ribbon. In general defects give rise to modes localized around the defect. They originate from the local change in force constants shifting the mode out of its unperturbed subband\cite{Savin2013}.

The relative motion of the two carbon atoms in each dimer only couples weakly to the motion of neighbouring dimers and to the phonons in the leads. This high-energy mode is shifted out of the entire phonon bandstructure. Meanwhile, the flowing electrical current passing through
these dimers introduce a small bias-dependent coupling via the self-energy ($\BPi[e]$) in Eq.~\ref{eq:WBAdouble}. In principle, one can tune the relative distance between different dimers by changing the ribbon width or the position of the
dehydrogenation. It is an ideal and clean system to study the
current-induced atomic dynamics, with some tunability since one may imagine doping or gating to shift the Fermi level, $E_F$, as well as changes in geometry such as varying the distance between dimers. 

In this study, the nanoribbon has a width of $7$ dimers corresponding to a C-C edge distance of 7.5 {\AA}.  The
lateral confinement introduces a direct semi-conducting band gap, giving rise to the gap in the electronic transmission for the perfect ribbon as shown in \Figref{fig:Setup} (b). 
We have introduced a broadening of the electronic states as in Christensen \etal\cite{Christensen2015} to mimic coupling to metallic electrodes. This in effect smoothen the transmission curves \Figref{fig:Setup} (b, dotted lines) akin to the experimental conductance\cite{Koch2012}. The introduction of the defects results mainly in a potential shift, but besides this does not impact the transmission dramatically, as seen in \Figref{fig:Setup} (b, full lines).

In order to characterize the phonons we show the phononic transmission in
\Figref{fig:Setup} (c). The phonon transmission shows a significant reduction of $\sim 50\%$ for $\hbar\omega$ above 25 meV. 
The fact that we see little difference between perfect and defected system at low phonon energy is expected when the wavelength is much larger than the defect.
We note that, ideally, the translation in the three spatial directions and rotation around the longitudinal direction of the GNR should lead to perfect $T=4$ at
$\hbar\omega$ equal exactly to zero for both pristine and defected structure. The deviation is due to our numerical neglect of long range elastic forces. However, while the low energy/long wavelength modes are important for heat transport, we are here concerned with  modes with a higher frequency above 25~meV, where the calculation is expected to be accurate.

\begin{figure}[t]
   \includegraphics[width=0.5\textwidth]{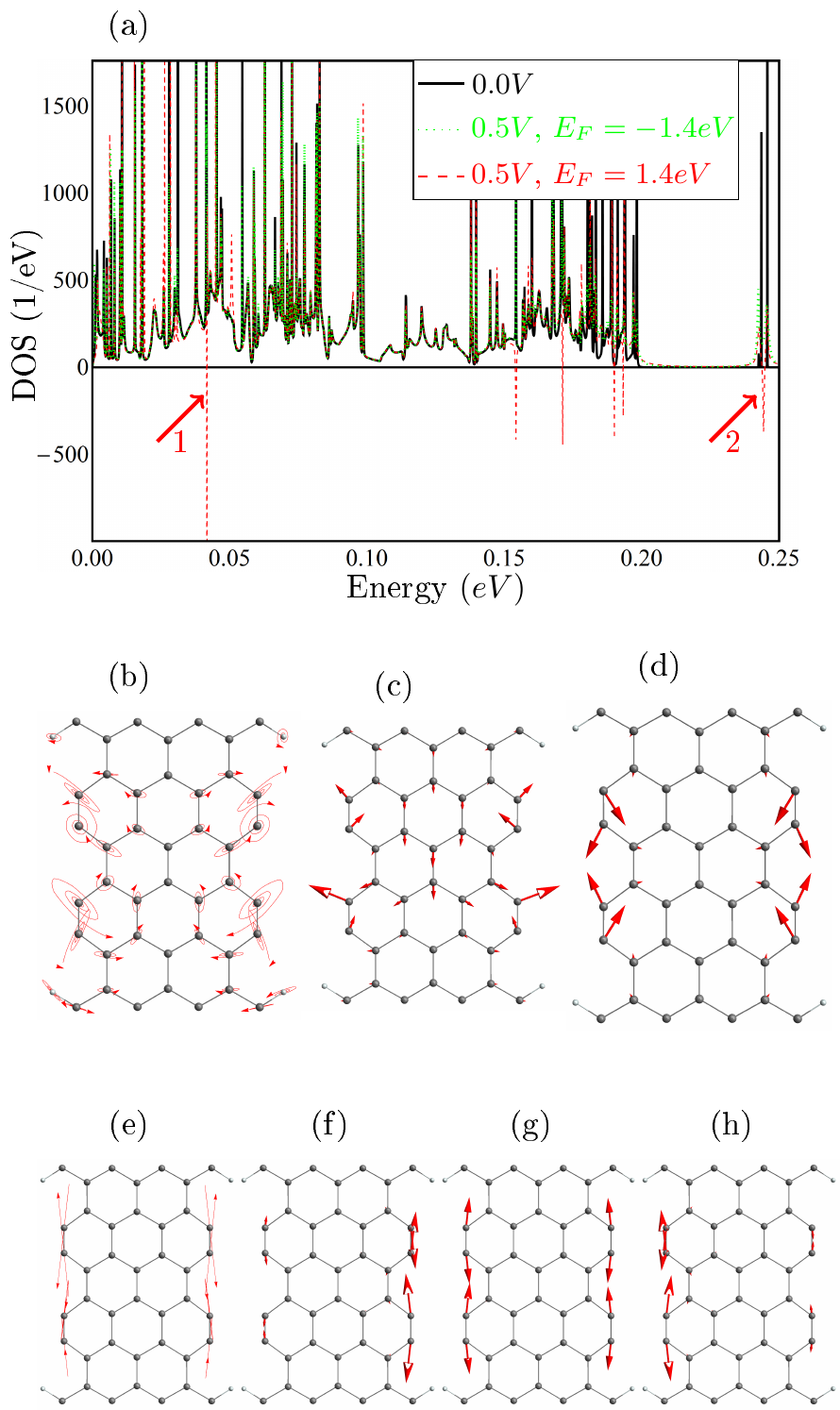}
   \caption{(a) The black solid line shows the phonon density of states
     excluding the self-energy due to the electronic degrees of freedom. The
     green dotted and the red dashed lines show the phonon DOS including the
     current induced forces for an applied bias of $0.5$ V,  shifting the Fermi
     energy to $E_F=-1.4$ eV and $E_F=1.4$ eV, respectively, corresponding to
     the vertical lines in \Figref{fig:Setup} (b). (b) the run-away mode giving
   rise to the dip in (a) indicted by arrow $1$ ($Q\sim 10^{-3}$), (c)-(d) the most important
 normal modes taking part in (b). (e) the run-away mode giving rise to the dip
 in (a) indicted by arrow $2$ ($Q\sim 0.5\cdot 10^{-3}$), (f)-(h) the normal modes taking part in (b). }
   \label{fig:DOS}
\end{figure}

The influence of the current-induced forces on the phonon self-energy depend on the underlying electronic properties. Thus, besides the unperturbed phonon DOS exclusive the current-induced forces (black solid line in \Figref{fig:DOS}
(a)), we also calculate the non-equilibrium DOS using \Eqref{eq:PHDOS}
shifting the Fermi energy away from the gap to $E_F=-1.4$~eV (green dotted
line) and $E_F=1.4$~eV (red dashed line), and applying a bias of $V_b=0.5$~V.
Comparing these results, we see that there are several run-away modes (negative peaks) for
$E_F=1.4$~eV, but not for $E_F=-1.4$~eV. 
In \Figref{fig:DOS} (b) and
(e) we show two typical run-away modes marked as $1$ and $2$ in \Figref{fig:DOS} (a). 
In (c)-(d) and (f)-(h), we also plot the "bare" normal modes (without coupling to electron and phonon baths) that give the largest
contribution to the two selected run-away modes. 
These run-away modes have in common that they are spatially localized, meaning a vanishing damping due to the coupling to the phonon reservoirs, 
which is a prerequisite for the run-away instability here. The driving of the motion by the current has to exceed this damping.
Both run-away modes involve mainly the dehydrogenated carbon dimers, but where mode $2$ lies outside the bulk phonon bands and is thus localized,  
the frequency of mode $1$ is well within the entire phonon band, illustrating that this is not a necessary requirement. In the case of $1$ the localization is due to a shift out of a ribbon phonon sub-band.
Most of the "bare" normal modes ((c)-(d), (f)-(h)) contributing to run-away
motion can be considered as a in or out of phase combination of different dimer
vibrations.

The two selected modes illustrate how the run-away modes can display circular motion in real-space or in abstract mode space. For the modes showing circular motion it is intuitively clear why these have been dubbed "water-wheel" modes\cite{DuMcTo.2009}, and that the direction of rotation is linked to the direction of current via (angular) momentum transfer\cite{ToDuDu.2011}. Mode $1$ is made up by two principal bare modes, while
mode $2$ does so in abstract mode space, and consists of mainly 3 bare modes. 
In both cases the nonconservative force pump energy into these modes when they oscillate around closed loops.  We note that anharmonic coupling will lead to additional damping which is not included in the harmonic approximation applied here, but we expect a lower anharmonic coupling to the mode well outside the bulk band ($2$) due to the frequency mismatch. The structure we consider has mirror symmetry in the lateral direction
perpendicular to the transport. The resulting motion of the run-away mode
respects this symmetry. But, the motion along the current direction is
asymmetric since the current breaks the symmetry.  This is more obvious for mode $1$ (cf \Figref{fig:DOS} (c)). Mode $2$ is strictly
localized in the center, but mode $1$ has weak coupling to the leads.
Consequently, mode $1$ shows larger asymmetry. We should mention that this asymmetry in the local heating already shows up before the run-away modes emerge\cite{Lu2015}.

\section{Conclusions}
\label{sec:conclusion}
In conclusion, we have studied the effect of current-induced forces on the
dynamics of dehydrogenated carbon dimers at the edges of graphene armchair
ribbon. These carbon dimers are weakly coupled to each other and the rest
carbon atoms, but interact with the electrical current. This
induces effective coupling between them, and the non-conservative and effective 
magnetic force become important in describing their dynamics. Using a
simplified eigen-mode analysis, we analyze how the carbon dimer motion is modified
by these forces at different Fermi level positions. The possibility of observing the atomic structure of the two-dimensional structures in microscopy, gating or doping, and atomic scale modification of graphene ribbon boundaries makes it an ideal candidate to study current-induced forces in nano-conductors, where interesting theoretical predictions are awaiting for experimental confirmation.

So far current-induced motion and desorption have been observed around edges in graphene sheets\cite{JiaX2009sci,EngelundM2010prl}. One signature of the non-conservative forces is, besides the asymmetry build into the momentum transfer, the highly non-linear heating of modes with bias\cite{Lu2010e}, which in principle could be observed around edges\cite{Islam2014}.

\section{Acknowledgements}
We acknowledge computer resources from the DCSC, and support from Center for
Nano-structured Graphene (Project DNRF58). J.T.L. acknowledges support from the
National Natural Science Foundation of China (Grants No. 11304107 and No.
61371015).


\begin{thebibliography}{33}%
\makeatletter
\providecommand \@ifxundefined [1]{%
 \@ifx{#1\undefined}
}%
\providecommand \@ifnum [1]{%
 \ifnum #1\expandafter \@firstoftwo
 \else \expandafter \@secondoftwo
 \fi
}%
\providecommand \@ifx [1]{%
 \ifx #1\expandafter \@firstoftwo
 \else \expandafter \@secondoftwo
 \fi
}%
\providecommand \natexlab [1]{#1}%
\providecommand \enquote  [1]{``#1''}%
\providecommand \bibnamefont  [1]{#1}%
\providecommand \bibfnamefont [1]{#1}%
\providecommand \citenamefont [1]{#1}%
\providecommand \href@noop [0]{\@secondoftwo}%
\providecommand \href [0]{\begingroup \@sanitize@url \@href}%
\providecommand \@href[1]{\@@startlink{#1}\@@href}%
\providecommand \@@href[1]{\endgroup#1\@@endlink}%
\providecommand \@sanitize@url [0]{\catcode `\\12\catcode `\$12\catcode
  `\&12\catcode `\#12\catcode `\^12\catcode `\_12\catcode `\%12\relax}%
\providecommand \@@startlink[1]{}%
\providecommand \@@endlink[0]{}%
\providecommand \url  [0]{\begingroup\@sanitize@url \@url }%
\providecommand \@url [1]{\endgroup\@href {#1}{\urlprefix }}%
\providecommand \urlprefix  [0]{URL }%
\providecommand \Eprint [0]{\href }%
\providecommand \doibase [0]{http://dx.doi.org/}%
\providecommand \selectlanguage [0]{\@gobble}%
\providecommand \bibinfo  [0]{\@secondoftwo}%
\providecommand \bibfield  [0]{\@secondoftwo}%
\providecommand \translation [1]{[#1]}%
\providecommand \BibitemOpen [0]{}%
\providecommand \bibitemStop [0]{}%
\providecommand \bibitemNoStop [0]{.\EOS\space}%
\providecommand \EOS [0]{\spacefactor3000\relax}%
\providecommand \BibitemShut  [1]{\csname bibitem#1\endcsname}%
\let\auto@bib@innerbib\@empty
\bibitem [{\citenamefont {Castro~Neto}\ \emph {et~al.}(2009)\citenamefont
  {Castro~Neto}, \citenamefont {Guinea}, \citenamefont {Peres}, \citenamefont
  {Novoselov},\ and\ \citenamefont {Geim}}]{Neto2009}%
  \BibitemOpen
  \bibfield  {author} {\bibinfo {author} {\bibfnamefont {A.~H.}\ \bibnamefont
  {Castro~Neto}}, \bibinfo {author} {\bibfnamefont {F.}~\bibnamefont {Guinea}},
  \bibinfo {author} {\bibfnamefont {N.~M.~R.}\ \bibnamefont {Peres}}, \bibinfo
  {author} {\bibfnamefont {K.~S.}\ \bibnamefont {Novoselov}}, \ and\ \bibinfo
  {author} {\bibfnamefont {A.~K.}\ \bibnamefont {Geim}},\ }\href {\doibase
  10.1103/RevModPhys.81.109} {\bibfield  {journal} {\bibinfo  {journal} {Rev.
  Mod. Phys.}\ }\textbf {\bibinfo {volume} {81}},\ \bibinfo {pages} {109}
  (\bibinfo {year} {2009})}\BibitemShut {NoStop}%
\bibitem [{\citenamefont {Cai}\ \emph {et~al.}(2010)\citenamefont {Cai},
  \citenamefont {Ruffieux}, \citenamefont {Jaafar}, \citenamefont {Bieri},
  \citenamefont {Braun}, \citenamefont {Blankenburg}, \citenamefont {Muoth},
  \citenamefont {Seitsonen}, \citenamefont {Saleh}, \citenamefont {Feng},
  \citenamefont {M\"{u}llen},\ and\ \citenamefont {Fasel}}]{Cai2010}%
  \BibitemOpen
  \bibfield  {author} {\bibinfo {author} {\bibfnamefont {J.}~\bibnamefont
  {Cai}}, \bibinfo {author} {\bibfnamefont {P.}~\bibnamefont {Ruffieux}},
  \bibinfo {author} {\bibfnamefont {R.}~\bibnamefont {Jaafar}}, \bibinfo
  {author} {\bibfnamefont {M.}~\bibnamefont {Bieri}}, \bibinfo {author}
  {\bibfnamefont {T.}~\bibnamefont {Braun}}, \bibinfo {author} {\bibfnamefont
  {S.}~\bibnamefont {Blankenburg}}, \bibinfo {author} {\bibfnamefont
  {M.}~\bibnamefont {Muoth}}, \bibinfo {author} {\bibfnamefont {A.~P.}\
  \bibnamefont {Seitsonen}}, \bibinfo {author} {\bibfnamefont {M.}~\bibnamefont
  {Saleh}}, \bibinfo {author} {\bibfnamefont {X.}~\bibnamefont {Feng}},
  \bibinfo {author} {\bibfnamefont {K.}~\bibnamefont {M\"{u}llen}}, \ and\
  \bibinfo {author} {\bibfnamefont {R.}~\bibnamefont {Fasel}},\ }\href
  {\doibase 10.1038/nature09211} {\bibfield  {journal} {\bibinfo  {journal}
  {Nature}\ }\textbf {\bibinfo {volume} {466}},\ \bibinfo {pages} {470}
  (\bibinfo {year} {2010})}\BibitemShut {NoStop}%
\bibitem [{\citenamefont {Cai}\ \emph {et~al.}(2014)\citenamefont {Cai},
  \citenamefont {Pignedoli}, \citenamefont {Talirz}, \citenamefont {Ruffieux},
  \citenamefont {S"ode}, \citenamefont {Liang}, \citenamefont {Meunier},
  \citenamefont {Berger}, \citenamefont {Li}, \citenamefont {Feng},
  \citenamefont {M"ullen},\ and\ \citenamefont {Fasel}}]{Cai2014}%
  \BibitemOpen
  \bibfield  {author} {\bibinfo {author} {\bibfnamefont {J.}~\bibnamefont
  {Cai}}, \bibinfo {author} {\bibfnamefont {C.~a.}\ \bibnamefont {Pignedoli}},
  \bibinfo {author} {\bibfnamefont {L.}~\bibnamefont {Talirz}}, \bibinfo
  {author} {\bibfnamefont {P.}~\bibnamefont {Ruffieux}}, \bibinfo {author}
  {\bibfnamefont {H.}~\bibnamefont {S"ode}}, \bibinfo {author} {\bibfnamefont
  {L.}~\bibnamefont {Liang}}, \bibinfo {author} {\bibfnamefont
  {V.}~\bibnamefont {Meunier}}, \bibinfo {author} {\bibfnamefont
  {R.}~\bibnamefont {Berger}}, \bibinfo {author} {\bibfnamefont
  {R.}~\bibnamefont {Li}}, \bibinfo {author} {\bibfnamefont {X.}~\bibnamefont
  {Feng}}, \bibinfo {author} {\bibfnamefont {K.}~\bibnamefont {M"ullen}}, \
  and\ \bibinfo {author} {\bibfnamefont {R.}~\bibnamefont {Fasel}},\ }\href
  {\doibase 10.1038/nnano.2014.184} {\bibfield  {journal} {\bibinfo  {journal}
  {Nat. Nanotechnol.}\ }\textbf {\bibinfo {volume} {9}},\ \bibinfo {pages}
  {896} (\bibinfo {year} {2014})}\BibitemShut {NoStop}%
\bibitem [{\citenamefont {Liu}\ \emph {et~al.}(2015)\citenamefont {Liu},
  \citenamefont {Li}, \citenamefont {Tan}, \citenamefont {Giannakopoulos},
  \citenamefont {Sanchez-Sanchez}, \citenamefont {Beljonne}, \citenamefont
  {Ruffieux}, \citenamefont {Fasel}, \citenamefont {Feng},\ and\ \citenamefont
  {M\"{u}llen}}]{Liu2015}%
  \BibitemOpen
  \bibfield  {author} {\bibinfo {author} {\bibfnamefont {J.}~\bibnamefont
  {Liu}}, \bibinfo {author} {\bibfnamefont {B.-W.}\ \bibnamefont {Li}},
  \bibinfo {author} {\bibfnamefont {Y.-Z.}\ \bibnamefont {Tan}}, \bibinfo
  {author} {\bibfnamefont {A.}~\bibnamefont {Giannakopoulos}}, \bibinfo
  {author} {\bibfnamefont {C.}~\bibnamefont {Sanchez-Sanchez}}, \bibinfo
  {author} {\bibfnamefont {D.}~\bibnamefont {Beljonne}}, \bibinfo {author}
  {\bibfnamefont {P.}~\bibnamefont {Ruffieux}}, \bibinfo {author}
  {\bibfnamefont {R.}~\bibnamefont {Fasel}}, \bibinfo {author} {\bibfnamefont
  {X.}~\bibnamefont {Feng}}, \ and\ \bibinfo {author} {\bibfnamefont
  {K.}~\bibnamefont {M\"{u}llen}},\ }\href {\doibase 10.1021/jacs.5b03017}
  {\bibfield  {journal} {\bibinfo  {journal} {J. Am. Chem. Soc.}\ ,\ \bibinfo
  {pages} {150504124851003}} (\bibinfo {year} {2015})}\BibitemShut {NoStop}%
\bibitem [{\citenamefont {Koch}\ \emph {et~al.}(2012)\citenamefont {Koch},
  \citenamefont {Ample}, \citenamefont {Joachim},\ and\ \citenamefont
  {Grill}}]{Koch2012}%
  \BibitemOpen
  \bibfield  {author} {\bibinfo {author} {\bibfnamefont {M.}~\bibnamefont
  {Koch}}, \bibinfo {author} {\bibfnamefont {F.}~\bibnamefont {Ample}},
  \bibinfo {author} {\bibfnamefont {C.}~\bibnamefont {Joachim}}, \ and\
  \bibinfo {author} {\bibfnamefont {L.}~\bibnamefont {Grill}},\ }\href
  {\doibase 10.1038/nnano.2012.169} {\bibfield  {journal} {\bibinfo  {journal}
  {Nat. Nanotechnol.}\ }\textbf {\bibinfo {volume} {7}},\ \bibinfo {pages}
  {713} (\bibinfo {year} {2012})}\BibitemShut {NoStop}%
\bibitem [{\citenamefont {Christensen}\ \emph {et~al.}(2015)\citenamefont
  {Christensen}, \citenamefont {Frederiksen},\ and\ \citenamefont
  {Brandbyge}}]{Christensen2015}%
  \BibitemOpen
  \bibfield  {author} {\bibinfo {author} {\bibfnamefont {R.~B.}\ \bibnamefont
  {Christensen}}, \bibinfo {author} {\bibfnamefont {T.}~\bibnamefont
  {Frederiksen}}, \ and\ \bibinfo {author} {\bibfnamefont {M.}~\bibnamefont
  {Brandbyge}},\ }\href {\doibase 10.1103/PhysRevB.91.075434} {\bibfield
  {journal} {\bibinfo  {journal} {Phys. Rev. B}\ }\textbf {\bibinfo {volume}
  {91}},\ \bibinfo {pages} {075434} (\bibinfo {year} {2015})},\ \Eprint
  {http://arxiv.org/abs/1501.0226} {arXiv:1501.0226} \BibitemShut {NoStop}%
\bibitem [{\citenamefont {Jia}\ \emph {et~al.}(2009)\citenamefont {Jia},
  \citenamefont {Hofmann}, \citenamefont {Meunier}, \citenamefont {Sumpter},
  \citenamefont {Campos-Delgado}, \citenamefont {Romo-Herrera}, \citenamefont
  {Son}, \citenamefont {Hsieh}, \citenamefont {Reina}, \citenamefont {Kong},
  \citenamefont {Terrones},\ and\ \citenamefont {Dresselhaus}}]{JiaX2009sci}%
  \BibitemOpen
  \bibfield  {author} {\bibinfo {author} {\bibfnamefont {X.}~\bibnamefont
  {Jia}}, \bibinfo {author} {\bibfnamefont {M.}~\bibnamefont {Hofmann}},
  \bibinfo {author} {\bibfnamefont {V.}~\bibnamefont {Meunier}}, \bibinfo
  {author} {\bibfnamefont {B.~G.}\ \bibnamefont {Sumpter}}, \bibinfo {author}
  {\bibfnamefont {J.}~\bibnamefont {Campos-Delgado}}, \bibinfo {author}
  {\bibfnamefont {J.~M.}\ \bibnamefont {Romo-Herrera}}, \bibinfo {author}
  {\bibfnamefont {H.}~\bibnamefont {Son}}, \bibinfo {author} {\bibfnamefont
  {Y.-P.}\ \bibnamefont {Hsieh}}, \bibinfo {author} {\bibfnamefont
  {A.}~\bibnamefont {Reina}}, \bibinfo {author} {\bibfnamefont
  {J.}~\bibnamefont {Kong}}, \bibinfo {author} {\bibfnamefont {M.}~\bibnamefont
  {Terrones}}, \ and\ \bibinfo {author} {\bibfnamefont {M.~S.}\ \bibnamefont
  {Dresselhaus}},\ }\href@noop {} {\bibfield  {journal} {\bibinfo  {journal}
  {Science}\ }\textbf {\bibinfo {volume} {323}},\ \bibinfo {pages} {1701}
  (\bibinfo {year} {2009})}\BibitemShut {NoStop}%
\bibitem [{\citenamefont {Engelund}\ \emph {et~al.}(2010)\citenamefont
  {Engelund}, \citenamefont {F\"urst}, \citenamefont {Jauho},\ and\
  \citenamefont {Brandbyge}}]{EngelundM2010prl}%
  \BibitemOpen
  \bibfield  {author} {\bibinfo {author} {\bibfnamefont {M.}~\bibnamefont
  {Engelund}}, \bibinfo {author} {\bibfnamefont {J.~A.}\ \bibnamefont
  {F\"urst}}, \bibinfo {author} {\bibfnamefont {A.~P.}\ \bibnamefont {Jauho}},
  \ and\ \bibinfo {author} {\bibfnamefont {M.}~\bibnamefont {Brandbyge}},\
  }\href@noop {} {\bibfield  {journal} {\bibinfo  {journal} {Phys. Rev. Lett.}\
  }\textbf {\bibinfo {volume} {104}},\ \bibinfo {pages} {036807} (\bibinfo
  {year} {2010})}\BibitemShut {NoStop}%
\bibitem [{\citenamefont {Gunst}\ \emph {et~al.}(2013)\citenamefont {Gunst},
  \citenamefont {L\"u}, \citenamefont {Hedeg{\aa}rd},\ and\ \citenamefont
  {Brandbyge}}]{Gunst2013}%
  \BibitemOpen
  \bibfield  {author} {\bibinfo {author} {\bibfnamefont {T.}~\bibnamefont
  {Gunst}}, \bibinfo {author} {\bibfnamefont {J.-T.}\ \bibnamefont {L\"u}},
  \bibinfo {author} {\bibfnamefont {P.}~\bibnamefont {Hedeg{\aa}rd}}, \ and\
  \bibinfo {author} {\bibfnamefont {M.}~\bibnamefont {Brandbyge}},\ }\href
  {\doibase 10.1103/PhysRevB.88.161401} {\bibfield  {journal} {\bibinfo
  {journal} {Phys. Rev. B}\ }\textbf {\bibinfo {volume} {88}},\ \bibinfo
  {pages} {161401} (\bibinfo {year} {2013})},\ \bibinfo {note}
  {00000}\BibitemShut {NoStop}%
\bibitem [{\citenamefont {Dundas}\ \emph {et~al.}(2009)\citenamefont {Dundas},
  \citenamefont {McEniry},\ and\ \citenamefont {Todorov}}]{DuMcTo.2009}%
  \BibitemOpen
  \bibfield  {author} {\bibinfo {author} {\bibfnamefont {D.}~\bibnamefont
  {Dundas}}, \bibinfo {author} {\bibfnamefont {E.~J.}\ \bibnamefont {McEniry}},
  \ and\ \bibinfo {author} {\bibfnamefont {T.~N.}\ \bibnamefont {Todorov}},\
  }\href {\doibase 10.1038/NNANO.2008.411} {\bibfield  {journal} {\bibinfo
  {journal} {Nature Nanotech.}\ }\textbf {\bibinfo {volume} {4}},\ \bibinfo
  {pages} {99} (\bibinfo {year} {2009})}\BibitemShut {NoStop}%
\bibitem [{\citenamefont {L{\"{u}}}\ \emph {et~al.}(2010)\citenamefont
  {L{\"{u}}}, \citenamefont {Brandbyge},\ and\ \citenamefont
  {Hedeg{\aa}rd}}]{Lu2010e}%
  \BibitemOpen
  \bibfield  {author} {\bibinfo {author} {\bibfnamefont {J.-T.}\ \bibnamefont
  {L{\"{u}}}}, \bibinfo {author} {\bibfnamefont {M.}~\bibnamefont {Brandbyge}},
  \ and\ \bibinfo {author} {\bibfnamefont {P.}~\bibnamefont {Hedeg{\aa}rd}},\
  }\href {\doibase 10.1021/nl904233u} {\bibfield  {journal} {\bibinfo
  {journal} {Nano letters}\ }\textbf {\bibinfo {volume} {10}},\ \bibinfo
  {pages} {1657} (\bibinfo {year} {2010})}\BibitemShut {NoStop}%
\bibitem [{\citenamefont {L\"u}\ \emph {et~al.}(2012)\citenamefont {L\"u},
  \citenamefont {Brandbyge}, \citenamefont {Hedeg\aa{}rd}, \citenamefont
  {Todorov},\ and\ \citenamefont {Dundas}}]{luprb12}%
  \BibitemOpen
  \bibfield  {author} {\bibinfo {author} {\bibfnamefont {J.-T.}\ \bibnamefont
  {L\"u}}, \bibinfo {author} {\bibfnamefont {M.}~\bibnamefont {Brandbyge}},
  \bibinfo {author} {\bibfnamefont {P.}~\bibnamefont {Hedeg\aa{}rd}}, \bibinfo
  {author} {\bibfnamefont {T.~N.}\ \bibnamefont {Todorov}}, \ and\ \bibinfo
  {author} {\bibfnamefont {D.}~\bibnamefont {Dundas}},\ }\href {\doibase
  10.1103/PhysRevB.85.245444} {\bibfield  {journal} {\bibinfo  {journal} {Phys.
  Rev. B}\ }\textbf {\bibinfo {volume} {85}},\ \bibinfo {pages} {245444}
  (\bibinfo {year} {2012})}\BibitemShut {NoStop}%
\bibitem [{\citenamefont {Dundas}\ \emph {et~al.}(2012)\citenamefont {Dundas},
  \citenamefont {Cunningham}, \citenamefont {Buchanan}, \citenamefont
  {Terasawa}, \citenamefont {Paxton},\ and\ \citenamefont
  {Todorov}}]{Dundas2012}%
  \BibitemOpen
  \bibfield  {author} {\bibinfo {author} {\bibfnamefont {D.}~\bibnamefont
  {Dundas}}, \bibinfo {author} {\bibfnamefont {B.}~\bibnamefont {Cunningham}},
  \bibinfo {author} {\bibfnamefont {C.}~\bibnamefont {Buchanan}}, \bibinfo
  {author} {\bibfnamefont {A.}~\bibnamefont {Terasawa}}, \bibinfo {author}
  {\bibfnamefont {A.~T.}\ \bibnamefont {Paxton}}, \ and\ \bibinfo {author}
  {\bibfnamefont {T.~N.}\ \bibnamefont {Todorov}},\ }\href
  {http://arxiv.org/abs/1209.1324} {\bibfield  {journal} {\bibinfo  {journal}
  {arXiv:1209.1324}\ } (\bibinfo {year} {2012})}\BibitemShut {NoStop}%
\bibitem [{\citenamefont {Cunningham}\ \emph {et~al.}(2014)\citenamefont
  {Cunningham}, \citenamefont {Todorov},\ and\ \citenamefont
  {Dundas}}]{Cunningham2014}%
  \BibitemOpen
  \bibfield  {author} {\bibinfo {author} {\bibfnamefont {B.}~\bibnamefont
  {Cunningham}}, \bibinfo {author} {\bibfnamefont {T.~N.}\ \bibnamefont
  {Todorov}}, \ and\ \bibinfo {author} {\bibfnamefont {D.}~\bibnamefont
  {Dundas}},\ }\href {\doibase 10.1103/PhysRevB.90.115430} {\bibfield
  {journal} {\bibinfo  {journal} {Phys. Rev. B}\ }\textbf {\bibinfo {volume}
  {90}},\ \bibinfo {pages} {115430} (\bibinfo {year} {2014})}\BibitemShut
  {NoStop}%
\bibitem [{\citenamefont {Todorov}\ \emph {et~al.}(2014)\citenamefont
  {Todorov}, \citenamefont {Dundas}, \citenamefont {L\"u}, \citenamefont
  {Brandbyge},\ and\ \citenamefont {Hedeg{\aa}rd}}]{Todorov2014}%
  \BibitemOpen
  \bibfield  {author} {\bibinfo {author} {\bibfnamefont {T.~N.}\ \bibnamefont
  {Todorov}}, \bibinfo {author} {\bibfnamefont {D.}~\bibnamefont {Dundas}},
  \bibinfo {author} {\bibfnamefont {J.-T.}\ \bibnamefont {L\"u}}, \bibinfo
  {author} {\bibfnamefont {M.}~\bibnamefont {Brandbyge}}, \ and\ \bibinfo
  {author} {\bibfnamefont {P.}~\bibnamefont {Hedeg{\aa}rd}},\ }\href {\doibase
  10.1088/0143-0807/35/6/065004} {\bibfield  {journal} {\bibinfo  {journal}
  {Eur. J. Phys.}\ }\textbf {\bibinfo {volume} {35}},\ \bibinfo {pages}
  {065004} (\bibinfo {year} {2014})}\BibitemShut {NoStop}%
\bibitem [{\citenamefont {Bustos-Mar\'un}\ \emph {et~al.}(2013)\citenamefont
  {Bustos-Mar\'un}, \citenamefont {Refael},\ and\ \citenamefont {von
  Oppen}}]{Bustos2013}%
  \BibitemOpen
  \bibfield  {author} {\bibinfo {author} {\bibfnamefont {R.}~\bibnamefont
  {Bustos-Mar\'un}}, \bibinfo {author} {\bibfnamefont {G.}~\bibnamefont
  {Refael}}, \ and\ \bibinfo {author} {\bibfnamefont {F.}~\bibnamefont {von
  Oppen}},\ }\href {\doibase 10.1103/PhysRevLett.111.060802} {\bibfield
  {journal} {\bibinfo  {journal} {Phys. Rev. Lett.}\ }\textbf {\bibinfo
  {volume} {111}},\ \bibinfo {pages} {060802} (\bibinfo {year}
  {2013})}\BibitemShut {NoStop}%
\bibitem [{\citenamefont {Bode}\ \emph {et~al.}(2011)\citenamefont {Bode},
  \citenamefont {Kusminskiy}, \citenamefont {Egger},\ and\ \citenamefont {von
  Oppen}}]{BoKuEgVo.2011}%
  \BibitemOpen
  \bibfield  {author} {\bibinfo {author} {\bibfnamefont {N.}~\bibnamefont
  {Bode}}, \bibinfo {author} {\bibfnamefont {S.~V.}\ \bibnamefont
  {Kusminskiy}}, \bibinfo {author} {\bibfnamefont {R.}~\bibnamefont {Egger}}, \
  and\ \bibinfo {author} {\bibfnamefont {F.}~\bibnamefont {von Oppen}},\ }\href
  {\doibase 10.1103/PhysRevLett.107.036804} {\bibfield  {journal} {\bibinfo
  {journal} {Phys. Rev. Lett.}\ }\textbf {\bibinfo {volume} {107}},\ \bibinfo
  {pages} {036804} (\bibinfo {year} {2011})}\BibitemShut {NoStop}%
\bibitem [{\citenamefont {Feynman}\ and\ \citenamefont
  {Vernon}(1963)}]{FEYNMAN1963}%
  \BibitemOpen
  \bibfield  {author} {\bibinfo {author} {\bibfnamefont {R.~P.}\ \bibnamefont
  {Feynman}}\ and\ \bibinfo {author} {\bibfnamefont {F.~L.}\ \bibnamefont
  {Vernon}},\ }\href@noop {} {\bibfield  {journal} {\bibinfo  {journal} {Ann.
  Phys.}\ }\textbf {\bibinfo {volume} {24}},\ \bibinfo {pages} {118 } (\bibinfo
  {year} {1963})}\BibitemShut {NoStop}%
\bibitem [{\citenamefont {Caldeira}\ and\ \citenamefont
  {Leggett}(1983)}]{CALE.1983}%
  \BibitemOpen
  \bibfield  {author} {\bibinfo {author} {\bibfnamefont {A.}~\bibnamefont
  {Caldeira}}\ and\ \bibinfo {author} {\bibfnamefont {A.}~\bibnamefont
  {Leggett}},\ }\href@noop {} {\bibfield  {journal} {\bibinfo  {journal}
  {Physica A}\ }\textbf {\bibinfo {volume} {121}},\ \bibinfo {pages} {587}
  (\bibinfo {year} {1983})}\BibitemShut {NoStop}%
\bibitem [{\citenamefont {Schmid}(1982)}]{SC.1982}%
  \BibitemOpen
  \bibfield  {author} {\bibinfo {author} {\bibfnamefont {A.}~\bibnamefont
  {Schmid}},\ }\href@noop {} {\bibfield  {journal} {\bibinfo  {journal} {J. Low
  Temp. Phys.}\ }\textbf {\bibinfo {volume} {49}},\ \bibinfo {pages} {609}
  (\bibinfo {year} {1982})}\BibitemShut {NoStop}%
\bibitem [{\citenamefont {Head-Gordon}\ and\ \citenamefont
  {Tully}(1995)}]{HETU.1995}%
  \BibitemOpen
  \bibfield  {author} {\bibinfo {author} {\bibfnamefont {M.}~\bibnamefont
  {Head-Gordon}}\ and\ \bibinfo {author} {\bibfnamefont {J.~C.}\ \bibnamefont
  {Tully}},\ }\href@noop {} {\bibfield  {journal} {\bibinfo  {journal} {J.
  Chem. Phys.}\ }\textbf {\bibinfo {volume} {103}},\ \bibinfo {pages} {10137}
  (\bibinfo {year} {1995})}\BibitemShut {NoStop}%
\bibitem [{\citenamefont {L\"u}\ \emph {et~al.}(2011)\citenamefont {L\"u},
  \citenamefont {Gunst}, \citenamefont {Hedeg{\aa}rd},\ and\ \citenamefont
  {Brandbyge}}]{Lu2011}%
  \BibitemOpen
  \bibfield  {author} {\bibinfo {author} {\bibfnamefont {J.-T.}\ \bibnamefont
  {L\"u}}, \bibinfo {author} {\bibfnamefont {T.}~\bibnamefont {Gunst}},
  \bibinfo {author} {\bibfnamefont {P.}~\bibnamefont {Hedeg{\aa}rd}}, \ and\
  \bibinfo {author} {\bibfnamefont {M.}~\bibnamefont {Brandbyge}},\ }\href
  {\doibase 10.3762/bjnano.2.90} {\bibfield  {journal} {\bibinfo  {journal}
  {Beilstein J. Nanotechnol.}\ }\textbf {\bibinfo {volume} {2}},\ \bibinfo
  {pages} {814} (\bibinfo {year} {2011})}\BibitemShut {NoStop}%
\bibitem [{\citenamefont {Brandbyge}\ and\ \citenamefont
  {Hedeg{\aa}rd}(1994)}]{BrHe.1994a}%
  \BibitemOpen
  \bibfield  {author} {\bibinfo {author} {\bibfnamefont {M.}~\bibnamefont
  {Brandbyge}}\ and\ \bibinfo {author} {\bibfnamefont {P.}~\bibnamefont
  {Hedeg{\aa}rd}},\ }\href@noop {} {\bibfield  {journal} {\bibinfo  {journal}
  {Phys. Rev. Lett.}\ }\textbf {\bibinfo {volume} {72}},\ \bibinfo {pages}
  {2919} (\bibinfo {year} {1994})}\BibitemShut {NoStop}%
\bibitem [{\citenamefont {Bode}\ \emph {et~al.}(2012)\citenamefont {Bode},
  \citenamefont {Kusminskiy}, \citenamefont {Egger},\ and\ \citenamefont {von
  Oppen}}]{BoKuEg.2012}%
  \BibitemOpen
  \bibfield  {author} {\bibinfo {author} {\bibfnamefont {N.}~\bibnamefont
  {Bode}}, \bibinfo {author} {\bibfnamefont {S.~V.}\ \bibnamefont
  {Kusminskiy}}, \bibinfo {author} {\bibfnamefont {R.}~\bibnamefont {Egger}}, \
  and\ \bibinfo {author} {\bibfnamefont {F.}~\bibnamefont {von Oppen}},\ }\href
  {\doibase 10.3762/bjnano.3.15} {\bibfield  {journal} {\bibinfo  {journal}
  {Beilstein J. Nanotechnol.}\ }\textbf {\bibinfo {volume} {3}},\ \bibinfo
  {pages} {144} (\bibinfo {year} {2012})}\BibitemShut {NoStop}%
\bibitem [{\citenamefont {Soler}\ \emph {et~al.}(2002)\citenamefont {Soler},
  \citenamefont {Artacho}, \citenamefont {Gale}, \citenamefont {Garcia},
  \citenamefont {Junquera}, \citenamefont {Ordejon},\ and\ \citenamefont
  {Sanchez-Portal}}]{Soler.02}%
  \BibitemOpen
  \bibfield  {author} {\bibinfo {author} {\bibfnamefont {J.~M.}\ \bibnamefont
  {Soler}}, \bibinfo {author} {\bibfnamefont {E.}~\bibnamefont {Artacho}},
  \bibinfo {author} {\bibfnamefont {J.~D.}\ \bibnamefont {Gale}}, \bibinfo
  {author} {\bibfnamefont {A.}~\bibnamefont {Garcia}}, \bibinfo {author}
  {\bibfnamefont {J.}~\bibnamefont {Junquera}}, \bibinfo {author}
  {\bibfnamefont {P.}~\bibnamefont {Ordejon}}, \ and\ \bibinfo {author}
  {\bibfnamefont {D.}~\bibnamefont {Sanchez-Portal}},\ }\href@noop {}
  {\bibfield  {journal} {\bibinfo  {journal} {J. Phys.:Condens. Matter}\
  }\textbf {\bibinfo {volume} {14}},\ \bibinfo {pages} {2745} (\bibinfo {year}
  {2002})}\BibitemShut {NoStop}%
\bibitem [{\citenamefont {Brandbyge}\ \emph {et~al.}(2002)\citenamefont
  {Brandbyge}, \citenamefont {Mozos}, \citenamefont {Ordejon}, \citenamefont
  {Taylor},\ and\ \citenamefont {Stokbro}}]{BrMoOr.2002}%
  \BibitemOpen
  \bibfield  {author} {\bibinfo {author} {\bibfnamefont {M.}~\bibnamefont
  {Brandbyge}}, \bibinfo {author} {\bibfnamefont {J.-L.}\ \bibnamefont
  {Mozos}}, \bibinfo {author} {\bibfnamefont {P.}~\bibnamefont {Ordejon}},
  \bibinfo {author} {\bibfnamefont {J.}~\bibnamefont {Taylor}}, \ and\ \bibinfo
  {author} {\bibfnamefont {K.}~\bibnamefont {Stokbro}},\ }\href@noop {}
  {\bibfield  {journal} {\bibinfo  {journal} {Phys. Rev. B}\ }\textbf {\bibinfo
  {volume} {65}},\ \bibinfo {pages} {165401} (\bibinfo {year}
  {2002})}\BibitemShut {NoStop}%
\bibitem [{\citenamefont {Frederiksen}\ \emph {et~al.}(2007)\citenamefont
  {Frederiksen}, \citenamefont {Paulsson}, \citenamefont {Brandbyge},\ and\
  \citenamefont {Jauho}}]{FrPaBr.2007}%
  \BibitemOpen
  \bibfield  {author} {\bibinfo {author} {\bibfnamefont {T.}~\bibnamefont
  {Frederiksen}}, \bibinfo {author} {\bibfnamefont {M.}~\bibnamefont
  {Paulsson}}, \bibinfo {author} {\bibfnamefont {M.}~\bibnamefont {Brandbyge}},
  \ and\ \bibinfo {author} {\bibfnamefont {A.-P.}\ \bibnamefont {Jauho}},\
  }\href {\doibase 10.1103/PhysRevB.75.205413} {\bibfield  {journal} {\bibinfo
  {journal} {Phys. Rev. B}\ }\textbf {\bibinfo {volume} {75}},\ \bibinfo
  {pages} {205413} (\bibinfo {year} {2007})}\BibitemShut {NoStop}%
\bibitem [{\citenamefont {Wang}\ \emph {et~al.}(2010)\citenamefont {Wang},
  \citenamefont {Kr{\"{o}}ger}, \citenamefont {Berndt}, \citenamefont
  {V{\'{a}}zquez}, \citenamefont {Brandbyge},\ and\ \citenamefont
  {Paulsson}}]{Wang2010}%
  \BibitemOpen
  \bibfield  {author} {\bibinfo {author} {\bibfnamefont {Y.~F.}\ \bibnamefont
  {Wang}}, \bibinfo {author} {\bibfnamefont {J.}~\bibnamefont {Kr{\"{o}}ger}},
  \bibinfo {author} {\bibfnamefont {R.}~\bibnamefont {Berndt}}, \bibinfo
  {author} {\bibfnamefont {H.}~\bibnamefont {V{\'{a}}zquez}}, \bibinfo {author}
  {\bibfnamefont {M.}~\bibnamefont {Brandbyge}}, \ and\ \bibinfo {author}
  {\bibfnamefont {M.}~\bibnamefont {Paulsson}},\ }\href {\doibase
  10.1103/PhysRevLett.104.176802} {\bibfield  {journal} {\bibinfo  {journal}
  {Physical Review Letters}\ }\textbf {\bibinfo {volume} {104}},\ \bibinfo
  {pages} {176802} (\bibinfo {year} {2010})}\BibitemShut {NoStop}%
\bibitem [{\citenamefont {L\"u}\ \emph {et~al.}(2015)\citenamefont {L\"u},
  \citenamefont {Christensen}, \citenamefont {Wang}, \citenamefont
  {Hedeg{\aa}rd},\ and\ \citenamefont {Brandbyge}}]{Lu2015}%
  \BibitemOpen
  \bibfield  {author} {\bibinfo {author} {\bibfnamefont {J.-T.}\ \bibnamefont
  {L\"u}}, \bibinfo {author} {\bibfnamefont {R.~B.}\ \bibnamefont
  {Christensen}}, \bibinfo {author} {\bibfnamefont {J.-S.}\ \bibnamefont
  {Wang}}, \bibinfo {author} {\bibfnamefont {P.}~\bibnamefont {Hedeg{\aa}rd}},
  \ and\ \bibinfo {author} {\bibfnamefont {M.}~\bibnamefont {Brandbyge}},\
  }\href {\doibase 10.1103/PhysRevLett.114.096801} {\bibfield  {journal}
  {\bibinfo  {journal} {Phys. Rev. Lett.}\ }\textbf {\bibinfo {volume} {114}},\
  \bibinfo {pages} {096801} (\bibinfo {year} {2015})}\BibitemShut {NoStop}%
\bibitem [{\citenamefont {Todorov}\ \emph
  {et~al.}(2011{\natexlab{a}})\citenamefont {Todorov}, \citenamefont {Dundas},
  \citenamefont {Paxton},\ and\ \citenamefont {Horsfield}}]{Todorov2011}%
  \BibitemOpen
  \bibfield  {author} {\bibinfo {author} {\bibfnamefont {T.~N.}\ \bibnamefont
  {Todorov}}, \bibinfo {author} {\bibfnamefont {D.}~\bibnamefont {Dundas}},
  \bibinfo {author} {\bibfnamefont {A.~T.}\ \bibnamefont {Paxton}}, \ and\
  \bibinfo {author} {\bibfnamefont {A.~P.}\ \bibnamefont {Horsfield}},\ }\href
  {\doibase 10.3762/bjnano.2.79} {\bibfield  {journal} {\bibinfo  {journal}
  {Beilstein Journal of Nanotechnology}\ }\textbf {\bibinfo {volume} {2}},\
  \bibinfo {pages} {727} (\bibinfo {year} {2011}{\natexlab{a}})}\BibitemShut
  {NoStop}%
\bibitem [{\citenamefont {Savin}\ and\ \citenamefont
  {Kivshar}(2013)}]{Savin2013}%
  \BibitemOpen
  \bibfield  {author} {\bibinfo {author} {\bibfnamefont {A.~V.}\ \bibnamefont
  {Savin}}\ and\ \bibinfo {author} {\bibfnamefont {Y.~S.}\ \bibnamefont
  {Kivshar}},\ }\href {\doibase 10.1103/PhysRevB.88.125417} {\bibfield
  {journal} {\bibinfo  {journal} {Physical Review B - Condensed Matter and
  Materials Physics}\ }\textbf {\bibinfo {volume} {88}},\ \bibinfo {pages} {1}
  (\bibinfo {year} {2013})},\ \Eprint {http://arxiv.org/abs/arXiv:1304.7702v1}
  {arXiv:arXiv:1304.7702v1} \BibitemShut {NoStop}%
\bibitem [{\citenamefont {Todorov}\ \emph
  {et~al.}(2011{\natexlab{b}})\citenamefont {Todorov}, \citenamefont {Dundas},
  \citenamefont {Paxton},\ and\ \citenamefont {Horsfield}}]{ToDuDu.2011}%
  \BibitemOpen
  \bibfield  {author} {\bibinfo {author} {\bibfnamefont {T.~N.}\ \bibnamefont
  {Todorov}}, \bibinfo {author} {\bibfnamefont {D.}~\bibnamefont {Dundas}},
  \bibinfo {author} {\bibfnamefont {A.~T.}\ \bibnamefont {Paxton}}, \ and\
  \bibinfo {author} {\bibfnamefont {A.~P.}\ \bibnamefont {Horsfield}},\ }\href
  {\doibase 10.3762/bjnano.2.79} {\bibfield  {journal} {\bibinfo  {journal}
  {Beilstein J. Nanotechnol.}\ }\textbf {\bibinfo {volume} {2}},\ \bibinfo
  {pages} {727} (\bibinfo {year} {2011}{\natexlab{b}})}\BibitemShut {NoStop}%
\bibitem [{\citenamefont {Islam}\ \emph {et~al.}(2014)\citenamefont {Islam},
  \citenamefont {Tamakawa}, \citenamefont {Tanaka}, \citenamefont {Makino},\
  and\ \citenamefont {Hashimoto}}]{Islam2014}%
  \BibitemOpen
  \bibfield  {author} {\bibinfo {author} {\bibfnamefont {M.~S.}\ \bibnamefont
  {Islam}}, \bibinfo {author} {\bibfnamefont {D.}~\bibnamefont {Tamakawa}},
  \bibinfo {author} {\bibfnamefont {S.}~\bibnamefont {Tanaka}}, \bibinfo
  {author} {\bibfnamefont {T.}~\bibnamefont {Makino}}, \ and\ \bibinfo {author}
  {\bibfnamefont {A.}~\bibnamefont {Hashimoto}},\ }\href {\doibase
  10.1016/j.carbon.2014.06.023} {\bibfield  {journal} {\bibinfo  {journal}
  {Carbon}\ }\textbf {\bibinfo {volume} {77}},\ \bibinfo {pages} {1073}
  (\bibinfo {year} {2014})}\BibitemShut {NoStop}%
\end{thebibliography}
%

\end{document}